\documentclass[reprint,amsmath,amssymb,aps,twocolumn]{revtex4}
\usepackage{graphicx}
\usepackage{dcolumn}
\usepackage{gensymb}
\usepackage{bm}
\usepackage{gensymb}
\begin{document}
\title{High Quality factor micro-ring resonator for strong atom-light interactions using miniature atomic beams}
\author{Ali Eshaghian Dorche$^{1,\dagger}$, Bochao Wei$^{2,\dagger}$, Chandra Raman$^{2,*}$,Ali Adibi$^{1,*}$\footnotetext{$^\dagger$ These two authors contributed equally to this work}
\footnotetext{$^*$ Corresponding email: ali.adibi@ece.gatech.edu,
chandra.raman@physics.gatech.edu} }
\address{$^1$ School of Electrical and Computer Engineering, Georgia Institute of Technology, Atlanta, GA, USA}
\address{$^2$ School of Physics, Georgia Institute of Technology, Atlanta, GA, USA}

\date{\today}

\begin{abstract}
An integrated photonic platform is proposed for strong interactions between atomic beams and annealing-free high-quality-factor (Q) microresonators. We fabricated a thin-film, air-clad SiN microresonator with a loaded Q of $1.55\times10^6$ around the optical transition of $^{87}$Rb at $~780$ nm. This Q is achieved without annealing the devices at high temperatures, enabling future fully integrated platforms containing optoelectronic circuitry as well. The estimated single-photon Rabi frequency (2g) is ${2\boldsymbol{\pi}}\times$64 MHz at a height of 100 nm above the resonator. Our simulation result indicates that miniature atomic beams with a longitudinal speed of 0.2 m/s to 30 m/s will strongly interact with our resonator, allowing for the detection of single-atom transits and the realization of scalable single-atom photonic devices. Racetrack resonators with a similar Q can be used to detect thermal atomic beams with velocities around 300 m/s.
\end{abstract}
\maketitle
\section{Introduction}
Numerous applications in quantum optics, communications, and computing rely on strong interactions between single atoms and photons \cite{chang2018colloquium,reiserer2015cavity,o2013fiber}. Recently, atom-light interactions using nanophotonic devices \cite{thompson2013coupling,burgers2019clocked} and nanofibers \cite{kato2019observation,schneeweiss2017fiber} have been demonstrated in a number of laboratories. However, these setups require a substantial effort to cool, transport, and eventually trap the atoms near the dielectric surface of the nanophotonic system. Miniaturization and scalability are hindered by these additional requirements. Moreover, there is a mismatch of timescales between the nanoseconds required to strongly couple atoms and photons in microresonators and the milliseconds to seconds required for atomic cooling and trap loading. This limits the duty cycle of quantum devices based on atoms and photons. By contrast, laser slowed, or even in some cases, fast thermal atomic beams possess many of the prerequisites for quantum device fabrication \cite{li2019cascaded,li2020robust}. 

In this paper, we experimentally demonstrate a high-performance silicon nitride ($\mathrm{Si_3N_4}$ or in short, SiN) microring resonator and show, through simulations, that it is capable of strongly interacting with slow atomic beams during their microseconds-long transit time above the resonator.  In this feasibility study, we also show that racetrack versions of these resonators can be combined with thermal atomic beams, which substantially reduces the vacuum requirement and laser overhead and paves the way toward highly integrated quantum devices using atoms on chips.

Among different material platforms, stoichiometric low-pressure chemical-vapor-deposited SiN (LPCVD-$\mathrm{SiN}$) has garnered attention for integrated nanophotonics as a CMOS-compatible material with a wide transparency window ranging from short visible wavelengths ($450$ nm) to infrared (IR) \cite{baets2016silicon, munoz2017silicon}. This would allow the co-integration of photonic and microelectronic structures for system-level implementations \cite{rahim2017expanding}. High-quality-factor (high-Q) photonic resonators using SiN are demonstrated in the near-IR regime with Qs on the order of $40\times10^6$ \cite{ji2017ultra}. However, these record Qs are achieved in oxide-clad thick-film SiN resonators with a large radius. In most cases, high-temperature annealing is also used to reflow the etched structure, mitigate plasma damages, and further improve the Q of hitherto not-annealed microresonators \cite{li2013vertical, xuan2016high, brasch2014radiation}. However, such high-temperature annealing would degrade the underneath front-end-of-the-line (FEOL) circuitry, which is required in system-level implementations \cite{el2018annealing}. 

In our platform, the microresonator should allow a strong interaction of the atom flying above the resonator with the evanescent tail of the electric field of a resonant mode.  This requires a mode with low field confinement.  Furthermore, the microresonator needs to have a reasonably small radius, for the single atom cooperativity
parameter $C\sim Q/V$ increases with decreasing mode volume $V$. 

The requirements for efficient atom-photon interactions impose trade-offs among the degrees of freedom to achieve high Qs in microresonators. High-Q microresonators are usually achieved with a large radius, tight field confinement, oxide-clad. Our requirements along with the increased Rayleigh scattering at typical wavelengths of atomic transition \cite{gorodetsky2000rayleigh} highlight the challenges to achieve high-Q \cite{zhao2020visible}. In this paper, we demonstrate a LPCVD-SiN microring resonator with a loaded Q of $1.55\times10^6$ at near-visible wavelengths 
which is the record-high Q for the air-clad thin-film microrings with no chemical -mechanical polishing (CMP) at near-visible wavelengths, to the best of our knowledge.



A strong interaction between the cavity mode and the atom requires a thinner SiN film with air-cladding to expose the cavity mode. For this purpose, a SiN film with a thickness of $287$ nm is chosen. The width of the waveguide bent to form the microring resonator is chosen $w_f=5$ $\mu$m to minimize the surface roughness at the inner wall of the microresonator, while the second-order transverse electric (TE) mode (i.e., the electric field in the plane of resonator) with maximum electric field away from the outer wall would experience lower scattering compared to the fundamental TE mode. The outer radius of the ring is $R_o=35$ $\mu$m to ensure an appropriate quality of the resonant modes, i.e., avoiding significant curvature while increasing the coupling factor g, and having a large-enough spectral distance between different families of transverse resonant modes. The second-order radial TE mode has a higher Q than the first-order one as its peak of the electric field is farther away from the outer ring sidewall, thus lowering the scattering losses. For the $D_2$ atomic transition of $^{87}$Rb at the wavelength $\lambda = 780.24$ nm, fine-tuning of the resonance wavelength through thermal tuning is required.  This is compatible with device operation at an elevated temperature above 100\degree C to avoid atomic deposition.

\begin{figure}[t]
\centering
\includegraphics[trim = 0 50 200 0, width=\linewidth]{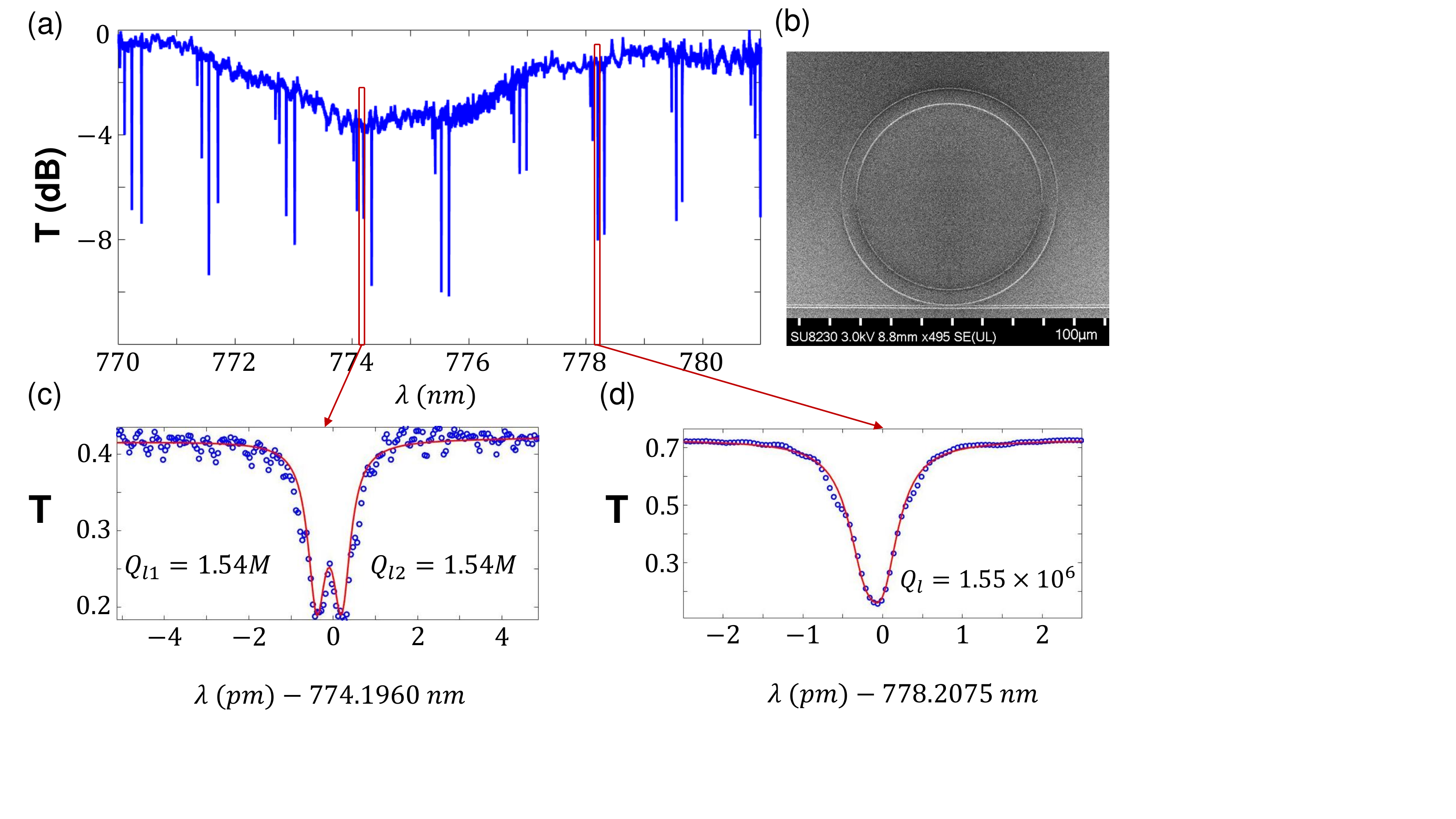}
\caption{ Optical characterization of the microring. (a) Normalized transmission through the waveguide, and (b) the scanning electron microscope (SEM) image of the fabricated microresonator with an outer radius of $35$ $\mu$m and a width of $5$ $\mu$m formed in a 287 nm-thick SiN on a $\mathrm{SiO_2}$ substrate. The microresonator is coupled to a bus waveguide with a gap distance of $100$ nm. In (a) families of modes spaced by $\sim 1$ nm can be seen.  Within each family, the spacing is large enough to avoid intermode coupling. The TE$_2$ family mode represents high-Q resonances as shown in the zoomed-in plots in (c) and (d). Loaded  Qs at both resonant wavelengths $\lambda = 774$ $(778.2)$ nm are above $1.5\times10^6$. Blue circles are experimental data and the solid red lines are fits to a double Lorentzian curve to account for the splitting of degenerate counter-clockwise resonant modes.}
\label{fig:characterization}
\end{figure}

The fabrication process starts with a planar polished oxidized silicon (Si) wafer, with $5$ $\mu$m thermal oxide that is formed by dry-wet-dry oxidation process, forming a $300$ nm-thick dry silicon dioxide (Si$O_2$) layer on top to ensure the high quality of the oxide region at the interface with the devices. The SiN layer for defining the waveguide and microresonator is deposited by LPCVD at a Tystar nitride furnace, with agent gases of dichlorosilane (DCS) and ammonia at $800 ^\circ C$. The stoichiometric SiN (i.e., $\mathrm{Si_3N_4}$) is deposited by adjusting the agents' ratio, leading to a refractive index around $2$ at $\lambda = 632$ nm. The deposition time is adjusted to reach a SiN thickness about $290$ nm. The sample is cleaned in two steps: first AMI solution (acetone, methanol, isopropanol), nitrogen blow dry, followed by BOE 6:1 (buffered oxide etchant) wet etching, running DI (deionized) water, and nitrogen blow dry. The sample is then baked on a hot plate to remove any moisture and prepare it for electron-beam resist coating. Flowable oxide (DOW corning FOx 16) is then spun on the SiN thin-film, followed by baking the sample on a hot plate. The devices are patterned using electron beam lithography (EBL) with current $I = 2 $ nA. The EBL parameters are optimized to minimize the edge roughness in the mask. After developing the sample in $25 \%$ Tetramethylammonium hydroxide (TMAH electronic grade) and nitrogen blow dry, the sample is etched in inductively coupled plasma (ICP) reactive ion etching (RIE) machine, with an optimized flow of oxygen, argon, and carbon tetrafluoride ($CF_4$) as the etching agents. The selectivity of SiN:FOx has been around one. The process is followed by short wet etching in BOE 6:1 and nitrogen blow dry.

To characterize the Q of the microresonators, the continuous-wave (CW) laser light from a tunable diode laser is passed through a polarization controller and input-coupled to the bus waveguide using a waveguide grating. The transmitted light is then detected at the output of the waveguide, where a multi-mode fiber collects light from an out-coupling grating and followed by a Si amplified detector. The CW laser is swept in the wavelength range $\lambda= 770 - 780.5$ nm with $0.05$ pm resolution.
The collected transmission from the detector is then fitted to two-Lorentzian curves, to take into account the intracavity coupling between counter-propagating modes. This facilitates the characterization of Qs associated with these modes.
Figure \ref{fig:characterization}(a) shows the transmission characteristics of the fabricated microring resonator with outer radius $R=$ 35 $\mu$m (see Fig. \ref{fig:characterization}(b)). 

The zoomed linear transmissions for two longitudinal resonant modes at $\lambda=774.2$ nm and $\lambda=778.2$ nm are depicted in Fig. \ref{fig:characterization}(c), and \ref{fig:characterization}(d), respectively, showing Qs larger than 1.5$\times 10^6$ for both modes.
Figure \ref{fig:characterization}(c) indicates the splitting between the counterclockwise and clockwise modes at $\lambda=774.2$ nm.
The splitting is proportional to the coupling between them, so for the resonant mode at $\lambda=778.20745$ nm, the coupling is almost negligible. 
This resonator can be temperature tuned to the $D_2$ resonance of the $^{87}$Rb atom at 780.24 nm.  Finite element simulations using COMSOL multiphysics and assuming a uniform temperature at the $\mathrm{SiO_2}$ and SiN layers indicate an $\approx 90 ^\circ C$ increase over the room temperature would be required. This number is roughly consistent with the formula $\frac{\partial \lambda}{\partial T} = \frac{\lambda}{n_g}\frac{\partial n_\text{eff}}{\partial T}$, with $n_g$ and $n_\text{eff}$ being the group index and the effective refractive index of the mode, respectively, T being the temperature in Kelvin (K), and using the thermo-optic coefficients (TOCs) $4\times10^{-5}$ $K^{-1}$ and $1.5\times 10^{-5}$ $K^{-1}$, respectively, for SiN and $\mathrm{SiO_2}$.

\begin{figure}[ht]
\centering
\includegraphics[width=\linewidth]{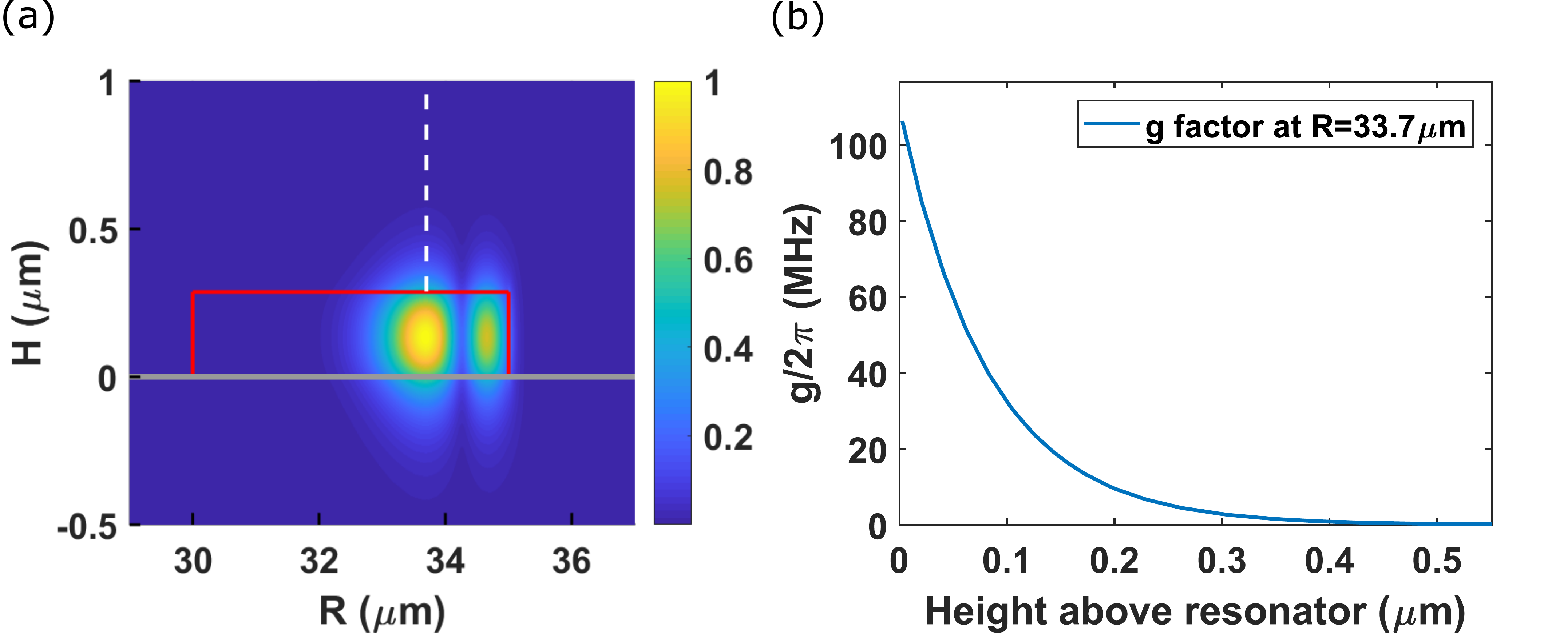}
\caption{High-Q SiN resonator for strong atom-light interaction. (a) Normalized electric field distribution at the cross-section of the microresonator for the mode $\mathrm{TE_{2,486}}$ with subscripts $2$ and $486$ showing the radial and azimuth mode numbers. The width of the SiN film bent to form the microring is $5$ $\mu$m, with a height of $287$ nm, and an outer radius of $35$ $\mu$m. Red lines show the boundary of the SiN resonator, and the gray line shows the boundary of the $\mathrm{SiO_2}$ substrate. (b) The atom-resonator coupling factor g(z) along the vertical dashed line in (a) at the lateral maximum of the resonator mode intensity at radius $R=33.7$ $\mu$m.}
\label{fig:first}
\end{figure}
The coupling between an atom and an optical mode is governed by the Jaynes–Cummings Hamiltonian \cite{reiserer2015cavity}:
\begin{equation}
        H=\hbar \omega_a\sigma^\dagger\sigma+\hbar \omega_c a^\dagger a+\hbar g(\vec{r})(\sigma a^\dagger+\sigma^\dagger a)
        \label{eq:hamiltonian}
\end{equation}
Here $\sigma = |g\rangle \langle e|$ is the atomic lowering operator, where $|g\rangle$ and $|e \rangle$ are the ground and the excited states of the atom, respectively;  $a$ is the photon annihilation operator; $\omega_a$ and $\omega_c$ are the atomic  and cavity angular resonance frequencies, respectively. The $g(\vec{r})$ factor characterizes the interaction strength between the atom and the cavity mode, which can be calculated as:
\begin{equation}
    g(\vec{r})=g_0\cdot \frac{\sqrt{\epsilon(\vec{r})}|E(\vec{r})|}{\max(\sqrt{\epsilon(\vec{r})}|E(\vec{r})|)}
\end{equation}
with $g_0=-d_{ge}\sqrt{\frac{\omega_c}{2\epsilon_0 \hbar V}}$. Here $d_{ge}$ is the effective dipole matrix element between ground and excited states; $V$ is the mode volume defined as $V=\int dr^3\frac{\epsilon(\vec{r})|E(\vec{r})|^2}{\max(\epsilon(\vec{r})|E(\vec{r})|^2)}$ \cite{englund2005controlling}, and $\epsilon(\vec{r})$ and $E(\vec{r})$ are material permittivity and electric field of the optical mode at location $\vec{r}$.
The electric field distribution of the $\mathrm{TE_{2,486}}$ mode is demonstrated in Figure \ref{fig:first}(a). The subscript numbers $2$ and $486$ indicate the radial and azimuth mode numbers, respectively.
The value of $g(\vec{r})$ at $R_0=33.7$ $\mu$m is shown in Figure \ref{fig:first}(b). It is noticed that $g(R_0,z)$ decreases very fast with $z$,  emphasizing the importance of air-cladding in resonator design.  Our resonator achieves a single-photon vacuum Rabi frequency $2g(R_0,z_0)=2\pi \times 64$ MHz around $z_0=100$ nm above the resonator while achieving a high Q with air-cladding. The single-atom cooperativity parameter is $C(R_0,z_0)=4g(R_0,z_0)^2/\kappa\gamma\approx2.7>1$. $\kappa$ and $\gamma$ here define the cavity and atomic energy decay rates.

\begin{figure}[t]
\centering
\includegraphics[width=\linewidth]{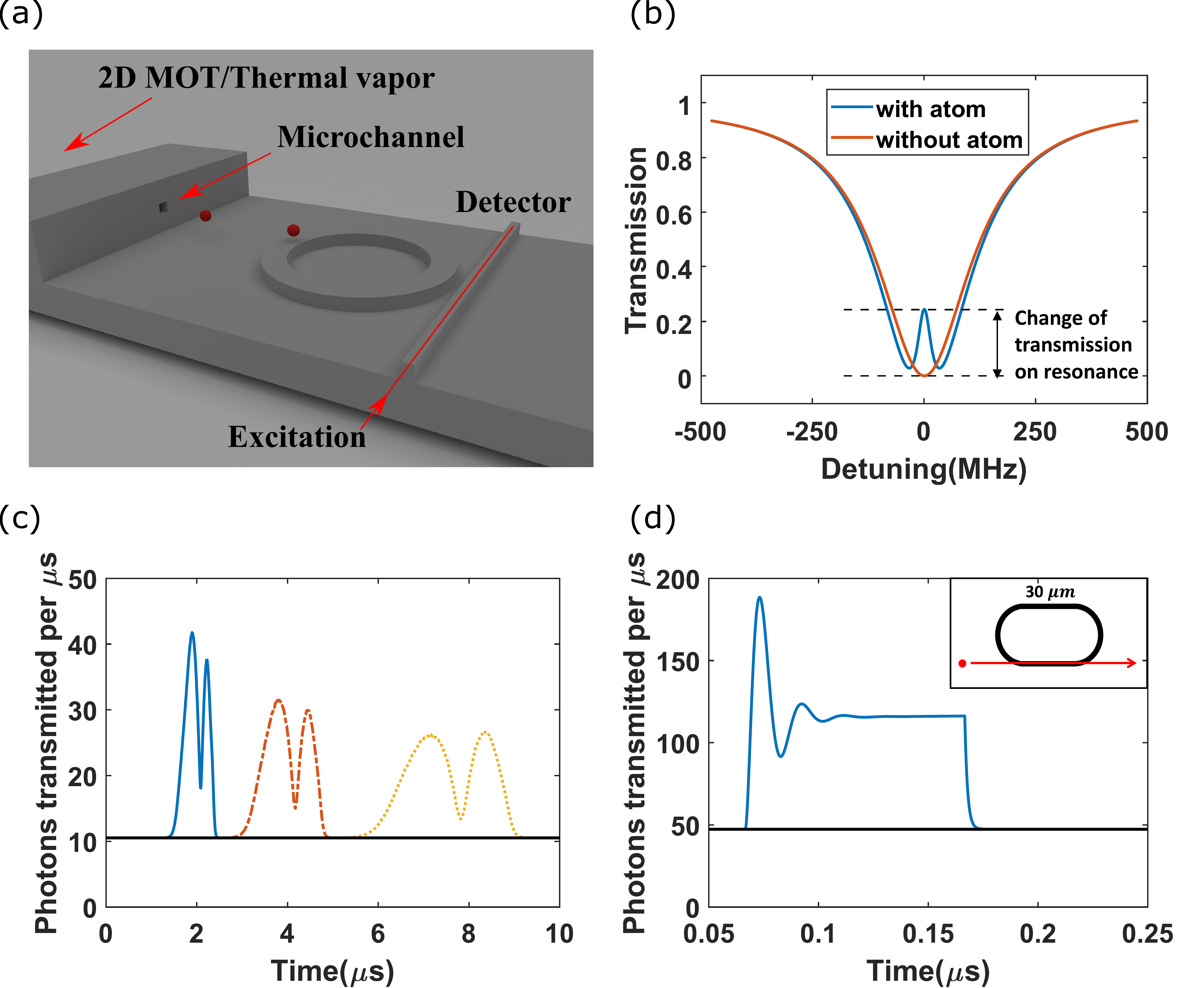}
\caption{(a) The schematic of the single-atom sensing platform. (b) Transmission spectra of excitation photons, with $g/2\pi=32$ MHz, loaded $Q=1.5\times10^6$, and 52.9 pW optical excitation power under the ideal critical coupling condition. The cavity is locked on resonance for atom sensing. (c) The detector signal when an atom flies above the resonator. The reference black line is the background signal from the imperfect critical coupling condition. Blue line: $v=3$ m/s, $h_0=150$ nm; Red dash: $v=1.5$ m/s, $h_0=190$ nm; Orange dot: $v=0.8$ m/s, $h_0=230$ nm. The excitation power is  52.9 pW. The vertical and horizontal velocities are set to be zero. $h_0$ is the vertical distance between atoms and the top resonator surface. (d) The detector signal when an atom flies above a racetrack resonator (shown in inset, with extinction ratio of 13 dB in the critical coupling regime), with velocity $v=300$ m/s, $h_0$ = $100$ nm, and excitation power of 231.8 pW. The resonator is locked to $w_0+kv_a$ to compensate the Doppler effect.}
\label{fig:simulation}
\end{figure}

Figure \ref{fig:simulation}(a) shows the concept of atom sensing using our high-Q resonator. Atoms from an atomic source, either a slow atomic beam from a 2D$^+$ magneto-optical trap (MOT) \cite{dieckmann1998two} or a beam of fast atoms from a thermal vapor as in Ref \cite{li2019cascaded}, pass through the microchannels, fly above the resonator, and strongly interact with it. Lithographically defined microchannels of 1-3 mm length help with atomic beam collimation and alignment with the resonator and will minimize excess contamination by alkali metals. The coupling waveguide is designed to bring the resonator into the critical coupling condition. 

To understand the dynamics of an atom interacting with the resonator we simulate the known Master equation \cite{reiserer2015cavity} for the density operator. We can apply the adiabatic approximation with slow atomic beams because the time scale for atomic transit (microseconds) over the resonator is much slower than the internal state evolution time (tens of nanoseconds) given ($(g,\ \gamma,\ \kappa)/2\pi\approx$ (32, 6, 256) MHz) and confirmed by full time-dependent calculations.  The spatial coordinate of the atom, $\vec{r}(t)$, is computed classically with the surface force, while the Master equation is solved using the instantaneous value of $g(\vec{r}(t))$. We use an approximate formula for the Casimir-Polder potential for Rb atoms (Ref \cite{thompson2013coupling} supplement)
$\mathrm{U}(z)=\frac{-C_{3}\frac{\lambda_{\text{eff}}}{2\pi}}{z^3\cdot(z+\frac{\lambda_{\text{eff}}}{2\pi})}$, where $C_3=2\pi\hbar\times1500$ $\mu$m$^3$, and $\lambda_{\text{eff}}=650$ nm for SiN. 

Figure \ref{fig:simulation}(b) shows the transmission of the cavity versus laser detuning for both an empty cavity and a cavity containing one stationary atom 100 nm above the resonator at $R=33.7 \mu$m.  The peak splitting is the well-known vacuum Rabi splitting that is proportional to $g(\vec{r})$ \cite{thompson1992observation}. If we set the excitation laser to zero-detuning and lock the resonator to the atomic transition, the presence of an atom will cause the transmission of the cavity to increase. With a constant excitation power, this will result in an increasing photon flux in the detector. For moving atoms, the change of the photon flux depends on $g(\vec{r}(t))$, and with single-photon counting modules as the detector, we can achieve the real-time monitoring of an atom transit. 


Sample transits for atoms with different velocities are shown in Figure \ref{fig:simulation}(c). The signals are displaced in time for a clearer view. The resonator size is the same as that in Figure \ref{fig:characterization} with a Q of $1.5\times10^6$.  The extinction ratio at the critical coupling condition is assumed to be 13 dB. The horizontal and vertical velocities of the atom are set to zero. The number of transmitted photons increases as an atom flies closer to the resonator surface, and the double peak in the signal results from the two nodes in the TE$_2$ field distribution (See Figure \ref{fig:first} (a)). Atoms with a broad range of longitudinal velocities can be detected. The atom's height and longitudinal velocity determine the signal intensity and duration, respectively. The faster the atom, the closer it can be to the resonator without crashing and the more intense (but shorter) signal it creates. With these constraints, we estimate that atoms with 0.2 m/s to 30 m/s longitudinal velocities can be detected without crashing on the resonator. The atom beams with such velocities can be generated from a 2D$^+$ MOT \cite{dieckmann1998two,ramirez2006multistage} and delivered via microchannels as shown schematically in Figure \ref{fig:simulation}(a). Interestingly, the ring configuration allows us to detect the same atom twice, and the time interval can be used to calculate the longitudinal velocity of the atom as well.

While the ratio $g/\kappa \approx0.12$ achieved is not yet in the strong coupling regime, the interaction between single atoms and photons is sufficiently strong that it can  enable quantum device applications.  For example, we envision that our platform could be used for a single-photon switch by single atoms, similar to what has been demonstrated using macroscopic resonators \cite{o2013fiber}. Using a 2D$^+$ MOT with $1.8\times10^{10}$ atoms/mm$^2$/s \cite{dieckmann1998two}, which has an average longitudinal velocity around 8 m/s, we estimate that with an excitation power of 3.4 photons/$\mu$s and a Q of $1.5\times10^6$, we can achieve on average, a 50\% contrast between "ON" and "OFF" states and a single photon routing rate around 10 kHz for each microchannel-resonator pair. The routing rate is much faster than those in the typical cold-atoms-dropping approach \cite{shomroni2014all}, which requires around one second to finish one cycle. If the resonator Q can be improved to $6\times10^6$, the contrast can be 85\%, which is competitive with the cold atom systems but with much higher repetition rates due to the absence of a long dead time. The switch can be activated by the entry of a single atom into the vicinity of the resonator and by using fast electronics to detect the rise of detected photons in the first 100 ns time bin. 

We also studied the potential use of racetrack resonators to detect thermal atomic beams generated by microchannel-collimation of a 100 \degree C Rb vapor \cite{li2019cascaded}. Atomic vapors have recently been proposed for cavity quantum electrodynamics (QED) applications \cite{alaeian2020cavity}.
The advantage of our approach is the ability to place atoms with lithographic precision at a discrete location above the resonator. We designed a racetrack resonator with a R = 20 $\mu$m ring section and 30 $\mu$m straight sections and align its straight sections to our microchannels (see Figure \ref{fig:simulation}(d) inset). In this way, the collimated thermal atoms with an average speed of 300 m/s will fly along the 30 $\mu$m straight section and strongly interact with the resonator for $\sim$ 100 ns. The Master equation simulation of a thermal atom transit without adiabatic approximation is shown in Figure \ref{fig:simulation}(d). The setup is similar to Figure \ref{fig:simulation}(a) and the Q of the racetrack resonator is assumed to be $1.5\times10^6$, with the transverse and vertical velocity set to be zero.  The initial peak is the Rabi oscillation resulting from an atom entering the regime with a strong evanescent field of the racetrack. If we monitor photons transmitted in 100 ns time bins and assume a 50\% detection efficiency, we should find the number of photons jumping from 2.5 to $\sim$ 6, which indicates an atom transit. We have fabricated racetrack resonators  with $\sim 7\times10^5$ loaded Q, and the optimization of the fabrication process for achieving Q $\sim1.5\times10^6$ is currently underway.

Our simulations reveal that miniature atomic beams are a promising candidate to achieve a strong interaction with the microresonator at the single-photon level. With the achieved Q and using multiple channel-resonator pairs together, one can imagine generating multi-photon states on chip in a simple and scalable architecture.  Without the requirement of atom trapping, ultra-high vacuum would not be needed. Thus, we believe that the platform reported here can set the stage for a new class of scalable quantum devices on chip using atoms and photons.

\bigskip
\noindent
{\textbf{Funding:}} This work was supported by the Air Force Office of Scentific Research (AFOSR) Grant no. FA9550-19-1-0228.  This work was performed in part at the Georgia Tech Institute for Electronics and Nanotechnology, a member of the National Nanotechnology Coordinated Infrastructure (NNCI), which is supported by the National Science Foundation (ECCS-1542174).

\bibliography{reference}

\begin{thebibliography}{25}
\expandafter\ifx\csname natexlab\endcsname\relax\def\natexlab#1{#1}\fi
\expandafter\ifx\csname bibnamefont\endcsname\relax
  \def\bibnamefont#1{#1}\fi
\expandafter\ifx\csname bibfnamefont\endcsname\relax
  \def\bibfnamefont#1{#1}\fi
\expandafter\ifx\csname citenamefont\endcsname\relax
  \def\citenamefont#1{#1}\fi
\expandafter\ifx\csname url\endcsname\relax
  \def\url#1{\texttt{#1}}\fi
\expandafter\ifx\csname urlprefix\endcsname\relax\def\urlprefix{URL }\fi
\providecommand{\bibinfo}[2]{#2}
\providecommand{\eprint}[2][]{\url{#2}}

\bibitem[{\citenamefont{Chang et~al.}(2018)\citenamefont{Chang, Douglas,
  Gonz{\'a}lez-Tudela, Hung, and Kimble}}]{chang2018colloquium}
\bibinfo{author}{\bibfnamefont{D.}~\bibnamefont{Chang}},
  \bibinfo{author}{\bibfnamefont{J.}~\bibnamefont{Douglas}},
  \bibinfo{author}{\bibfnamefont{A.}~\bibnamefont{Gonz{\'a}lez-Tudela}},
  \bibinfo{author}{\bibfnamefont{C.-L.} \bibnamefont{Hung}}, \bibnamefont{and}
  \bibinfo{author}{\bibfnamefont{H.}~\bibnamefont{Kimble}},
  \bibinfo{journal}{Reviews of Modern Physics} \textbf{\bibinfo{volume}{90}},
  \bibinfo{pages}{031002} (\bibinfo{year}{2018}).

\bibitem[{\citenamefont{Reiserer and Rempe}(2015)}]{reiserer2015cavity}
\bibinfo{author}{\bibfnamefont{A.}~\bibnamefont{Reiserer}} \bibnamefont{and}
  \bibinfo{author}{\bibfnamefont{G.}~\bibnamefont{Rempe}},
  \bibinfo{journal}{Reviews of Modern Physics} \textbf{\bibinfo{volume}{87}},
  \bibinfo{pages}{1379} (\bibinfo{year}{2015}).

\bibitem[{\citenamefont{O’Shea et~al.}(2013)\citenamefont{O’Shea, Junge,
  Volz, and Rauschenbeutel}}]{o2013fiber}
\bibinfo{author}{\bibfnamefont{D.}~\bibnamefont{O’Shea}},
  \bibinfo{author}{\bibfnamefont{C.}~\bibnamefont{Junge}},
  \bibinfo{author}{\bibfnamefont{J.}~\bibnamefont{Volz}}, \bibnamefont{and}
  \bibinfo{author}{\bibfnamefont{A.}~\bibnamefont{Rauschenbeutel}},
  \bibinfo{journal}{Physical review letters} \textbf{\bibinfo{volume}{111}},
  \bibinfo{pages}{193601} (\bibinfo{year}{2013}).

\bibitem[{\citenamefont{Thompson et~al.}(2013)\citenamefont{Thompson, Tiecke,
  de~Leon, Feist, Akimov, Gullans, Zibrov, Vuleti{\'c}, and
  Lukin}}]{thompson2013coupling}
\bibinfo{author}{\bibfnamefont{J.~D.} \bibnamefont{Thompson}},
  \bibinfo{author}{\bibfnamefont{T.}~\bibnamefont{Tiecke}},
  \bibinfo{author}{\bibfnamefont{N.~P.} \bibnamefont{de~Leon}},
  \bibinfo{author}{\bibfnamefont{J.}~\bibnamefont{Feist}},
  \bibinfo{author}{\bibfnamefont{A.}~\bibnamefont{Akimov}},
  \bibinfo{author}{\bibfnamefont{M.}~\bibnamefont{Gullans}},
  \bibinfo{author}{\bibfnamefont{A.~S.} \bibnamefont{Zibrov}},
  \bibinfo{author}{\bibfnamefont{V.}~\bibnamefont{Vuleti{\'c}}},
  \bibnamefont{and} \bibinfo{author}{\bibfnamefont{M.~D.} \bibnamefont{Lukin}},
  \bibinfo{journal}{Science} \textbf{\bibinfo{volume}{340}},
  \bibinfo{pages}{1202} (\bibinfo{year}{2013}).

\bibitem[{\citenamefont{Burgers et~al.}(2019)\citenamefont{Burgers, Peng,
  Muniz, McClung, Martin, and Kimble}}]{burgers2019clocked}
\bibinfo{author}{\bibfnamefont{A.~P.} \bibnamefont{Burgers}},
  \bibinfo{author}{\bibfnamefont{L.~S.} \bibnamefont{Peng}},
  \bibinfo{author}{\bibfnamefont{J.~A.} \bibnamefont{Muniz}},
  \bibinfo{author}{\bibfnamefont{A.~C.} \bibnamefont{McClung}},
  \bibinfo{author}{\bibfnamefont{M.~J.} \bibnamefont{Martin}},
  \bibnamefont{and} \bibinfo{author}{\bibfnamefont{H.~J.}
  \bibnamefont{Kimble}}, \bibinfo{journal}{Proceedings of the National Academy
  of Sciences} \textbf{\bibinfo{volume}{116}}, \bibinfo{pages}{456}
  (\bibinfo{year}{2019}).

\bibitem[{\citenamefont{Kato et~al.}(2019)\citenamefont{Kato, N{\'e}met, Senga,
  Mizukami, Huang, Parkins, and Aoki}}]{kato2019observation}
\bibinfo{author}{\bibfnamefont{S.}~\bibnamefont{Kato}},
  \bibinfo{author}{\bibfnamefont{N.}~\bibnamefont{N{\'e}met}},
  \bibinfo{author}{\bibfnamefont{K.}~\bibnamefont{Senga}},
  \bibinfo{author}{\bibfnamefont{S.}~\bibnamefont{Mizukami}},
  \bibinfo{author}{\bibfnamefont{X.}~\bibnamefont{Huang}},
  \bibinfo{author}{\bibfnamefont{S.}~\bibnamefont{Parkins}}, \bibnamefont{and}
  \bibinfo{author}{\bibfnamefont{T.}~\bibnamefont{Aoki}},
  \bibinfo{journal}{Nature communications} \textbf{\bibinfo{volume}{10}},
  \bibinfo{pages}{1} (\bibinfo{year}{2019}).

\bibitem[{\citenamefont{Schneeweiss et~al.}(2017)\citenamefont{Schneeweiss,
  Zeiger, Hoinkes, Rauschenbeutel, and Volz}}]{schneeweiss2017fiber}
\bibinfo{author}{\bibfnamefont{P.}~\bibnamefont{Schneeweiss}},
  \bibinfo{author}{\bibfnamefont{S.}~\bibnamefont{Zeiger}},
  \bibinfo{author}{\bibfnamefont{T.}~\bibnamefont{Hoinkes}},
  \bibinfo{author}{\bibfnamefont{A.}~\bibnamefont{Rauschenbeutel}},
  \bibnamefont{and} \bibinfo{author}{\bibfnamefont{J.}~\bibnamefont{Volz}},
  \bibinfo{journal}{Optics letters} \textbf{\bibinfo{volume}{42}},
  \bibinfo{pages}{85} (\bibinfo{year}{2017}).

\bibitem[{\citenamefont{Li et~al.}(2019)\citenamefont{Li, Chai, Wei, Yang,
  Daruwalla, Ayazi, and Raman}}]{li2019cascaded}
\bibinfo{author}{\bibfnamefont{C.}~\bibnamefont{Li}},
  \bibinfo{author}{\bibfnamefont{X.}~\bibnamefont{Chai}},
  \bibinfo{author}{\bibfnamefont{B.}~\bibnamefont{Wei}},
  \bibinfo{author}{\bibfnamefont{J.}~\bibnamefont{Yang}},
  \bibinfo{author}{\bibfnamefont{A.}~\bibnamefont{Daruwalla}},
  \bibinfo{author}{\bibfnamefont{F.}~\bibnamefont{Ayazi}}, \bibnamefont{and}
  \bibinfo{author}{\bibfnamefont{C.}~\bibnamefont{Raman}},
  \bibinfo{journal}{Nature communications} \textbf{\bibinfo{volume}{10}},
  \bibinfo{pages}{1} (\bibinfo{year}{2019}).

\bibitem[{\citenamefont{Li et~al.}(2020)\citenamefont{Li, Wei, Chai, Yang,
  Daruwalla, Ayazi, Raman et~al.}}]{li2020robust}
\bibinfo{author}{\bibfnamefont{C.}~\bibnamefont{Li}},
  \bibinfo{author}{\bibfnamefont{B.}~\bibnamefont{Wei}},
  \bibinfo{author}{\bibfnamefont{X.}~\bibnamefont{Chai}},
  \bibinfo{author}{\bibfnamefont{J.}~\bibnamefont{Yang}},
  \bibinfo{author}{\bibfnamefont{A.}~\bibnamefont{Daruwalla}},
  \bibinfo{author}{\bibfnamefont{F.}~\bibnamefont{Ayazi}},
  \bibinfo{author}{\bibfnamefont{C.}~\bibnamefont{Raman}},
  \bibnamefont{et~al.}, \bibinfo{journal}{Physical Review Research}
  \textbf{\bibinfo{volume}{2}}, \bibinfo{pages}{023239} (\bibinfo{year}{2020}).

\bibitem[{\citenamefont{Baets et~al.}(2016)\citenamefont{Baets, Subramanian,
  Clemmen, Kuyken, Bienstman, Le~Thomas, Roelkens, Van~Thourhout, Helin, and
  Severi}}]{baets2016silicon}
\bibinfo{author}{\bibfnamefont{R.}~\bibnamefont{Baets}},
  \bibinfo{author}{\bibfnamefont{A.~Z.} \bibnamefont{Subramanian}},
  \bibinfo{author}{\bibfnamefont{S.}~\bibnamefont{Clemmen}},
  \bibinfo{author}{\bibfnamefont{B.}~\bibnamefont{Kuyken}},
  \bibinfo{author}{\bibfnamefont{P.}~\bibnamefont{Bienstman}},
  \bibinfo{author}{\bibfnamefont{N.}~\bibnamefont{Le~Thomas}},
  \bibinfo{author}{\bibfnamefont{G.}~\bibnamefont{Roelkens}},
  \bibinfo{author}{\bibfnamefont{D.}~\bibnamefont{Van~Thourhout}},
  \bibinfo{author}{\bibfnamefont{P.}~\bibnamefont{Helin}}, \bibnamefont{and}
  \bibinfo{author}{\bibfnamefont{S.}~\bibnamefont{Severi}}, in
  \emph{\bibinfo{booktitle}{Optical Fiber Communication Conference}}
  (\bibinfo{organization}{Optical Society of America}, \bibinfo{year}{2016}),
  pp. \bibinfo{pages}{Th3J--1}.

\bibitem[{\citenamefont{Mu{\~n}oz et~al.}(2017)\citenamefont{Mu{\~n}oz,
  Mic{\'o}, Bru, Pastor, P{\'e}rez, Dom{\'e}nech, Fern{\'a}ndez, Ba{\~n}os,
  Gargallo, Alemany et~al.}}]{munoz2017silicon}
\bibinfo{author}{\bibfnamefont{P.}~\bibnamefont{Mu{\~n}oz}},
  \bibinfo{author}{\bibfnamefont{G.}~\bibnamefont{Mic{\'o}}},
  \bibinfo{author}{\bibfnamefont{L.~A.} \bibnamefont{Bru}},
  \bibinfo{author}{\bibfnamefont{D.}~\bibnamefont{Pastor}},
  \bibinfo{author}{\bibfnamefont{D.}~\bibnamefont{P{\'e}rez}},
  \bibinfo{author}{\bibfnamefont{J.~D.} \bibnamefont{Dom{\'e}nech}},
  \bibinfo{author}{\bibfnamefont{J.}~\bibnamefont{Fern{\'a}ndez}},
  \bibinfo{author}{\bibfnamefont{R.}~\bibnamefont{Ba{\~n}os}},
  \bibinfo{author}{\bibfnamefont{B.}~\bibnamefont{Gargallo}},
  \bibinfo{author}{\bibfnamefont{R.}~\bibnamefont{Alemany}},
  \bibnamefont{et~al.}, \bibinfo{journal}{Sensors}
  \textbf{\bibinfo{volume}{17}}, \bibinfo{pages}{2088} (\bibinfo{year}{2017}).

\bibitem[{\citenamefont{Rahim et~al.}(2017)\citenamefont{Rahim, Ryckeboer,
  Subramanian, Clemmen, Kuyken, Dhakal, Raza, Hermans, Muneeb, Dhoore
  et~al.}}]{rahim2017expanding}
\bibinfo{author}{\bibfnamefont{A.}~\bibnamefont{Rahim}},
  \bibinfo{author}{\bibfnamefont{E.}~\bibnamefont{Ryckeboer}},
  \bibinfo{author}{\bibfnamefont{A.~Z.} \bibnamefont{Subramanian}},
  \bibinfo{author}{\bibfnamefont{S.}~\bibnamefont{Clemmen}},
  \bibinfo{author}{\bibfnamefont{B.}~\bibnamefont{Kuyken}},
  \bibinfo{author}{\bibfnamefont{A.}~\bibnamefont{Dhakal}},
  \bibinfo{author}{\bibfnamefont{A.}~\bibnamefont{Raza}},
  \bibinfo{author}{\bibfnamefont{A.}~\bibnamefont{Hermans}},
  \bibinfo{author}{\bibfnamefont{M.}~\bibnamefont{Muneeb}},
  \bibinfo{author}{\bibfnamefont{S.}~\bibnamefont{Dhoore}},
  \bibnamefont{et~al.}, \bibinfo{journal}{Journal of lightwave technology}
  \textbf{\bibinfo{volume}{35}}, \bibinfo{pages}{639} (\bibinfo{year}{2017}).

\bibitem[{\citenamefont{Ji et~al.}(2017)\citenamefont{Ji, Barbosa, Roberts,
  Dutt, Cardenas, Okawachi, Bryant, Gaeta, and Lipson}}]{ji2017ultra}
\bibinfo{author}{\bibfnamefont{X.}~\bibnamefont{Ji}},
  \bibinfo{author}{\bibfnamefont{F.~A.} \bibnamefont{Barbosa}},
  \bibinfo{author}{\bibfnamefont{S.~P.} \bibnamefont{Roberts}},
  \bibinfo{author}{\bibfnamefont{A.}~\bibnamefont{Dutt}},
  \bibinfo{author}{\bibfnamefont{J.}~\bibnamefont{Cardenas}},
  \bibinfo{author}{\bibfnamefont{Y.}~\bibnamefont{Okawachi}},
  \bibinfo{author}{\bibfnamefont{A.}~\bibnamefont{Bryant}},
  \bibinfo{author}{\bibfnamefont{A.~L.} \bibnamefont{Gaeta}}, \bibnamefont{and}
  \bibinfo{author}{\bibfnamefont{M.}~\bibnamefont{Lipson}},
  \bibinfo{journal}{Optica} \textbf{\bibinfo{volume}{4}}, \bibinfo{pages}{619}
  (\bibinfo{year}{2017}).

\bibitem[{\citenamefont{Li et~al.}(2013)\citenamefont{Li, Eftekhar, Sodagar,
  Xia, Atabaki, and Adibi}}]{li2013vertical}
\bibinfo{author}{\bibfnamefont{Q.}~\bibnamefont{Li}},
  \bibinfo{author}{\bibfnamefont{A.~A.} \bibnamefont{Eftekhar}},
  \bibinfo{author}{\bibfnamefont{M.}~\bibnamefont{Sodagar}},
  \bibinfo{author}{\bibfnamefont{Z.}~\bibnamefont{Xia}},
  \bibinfo{author}{\bibfnamefont{A.~H.} \bibnamefont{Atabaki}},
  \bibnamefont{and} \bibinfo{author}{\bibfnamefont{A.}~\bibnamefont{Adibi}},
  \bibinfo{journal}{Optics express} \textbf{\bibinfo{volume}{21}},
  \bibinfo{pages}{18236} (\bibinfo{year}{2013}).

\bibitem[{\citenamefont{Xuan et~al.}(2016)\citenamefont{Xuan, Liu, Varghese,
  Metcalf, Xue, Wang, Han, Jaramillo-Villegas, Al~Noman, Wang
  et~al.}}]{xuan2016high}
\bibinfo{author}{\bibfnamefont{Y.}~\bibnamefont{Xuan}},
  \bibinfo{author}{\bibfnamefont{Y.}~\bibnamefont{Liu}},
  \bibinfo{author}{\bibfnamefont{L.~T.} \bibnamefont{Varghese}},
  \bibinfo{author}{\bibfnamefont{A.~J.} \bibnamefont{Metcalf}},
  \bibinfo{author}{\bibfnamefont{X.}~\bibnamefont{Xue}},
  \bibinfo{author}{\bibfnamefont{P.-H.} \bibnamefont{Wang}},
  \bibinfo{author}{\bibfnamefont{K.}~\bibnamefont{Han}},
  \bibinfo{author}{\bibfnamefont{J.~A.} \bibnamefont{Jaramillo-Villegas}},
  \bibinfo{author}{\bibfnamefont{A.}~\bibnamefont{Al~Noman}},
  \bibinfo{author}{\bibfnamefont{C.}~\bibnamefont{Wang}}, \bibnamefont{et~al.},
  \bibinfo{journal}{Optica} \textbf{\bibinfo{volume}{3}}, \bibinfo{pages}{1171}
  (\bibinfo{year}{2016}).

\bibitem[{\citenamefont{Brasch et~al.}(2014)\citenamefont{Brasch, Chen,
  Schiller, and Kippenberg}}]{brasch2014radiation}
\bibinfo{author}{\bibfnamefont{V.}~\bibnamefont{Brasch}},
  \bibinfo{author}{\bibfnamefont{Q.-F.} \bibnamefont{Chen}},
  \bibinfo{author}{\bibfnamefont{S.}~\bibnamefont{Schiller}}, \bibnamefont{and}
  \bibinfo{author}{\bibfnamefont{T.~J.} \bibnamefont{Kippenberg}},
  \bibinfo{journal}{Optics express} \textbf{\bibinfo{volume}{22}},
  \bibinfo{pages}{30786} (\bibinfo{year}{2014}).

\bibitem[{\citenamefont{El~Dirani et~al.}(2018)\citenamefont{El~Dirani, Kamel,
  Casale, Kerdiles, Monat, Letartre, Pu, Oxenl{\o}we, Yvind, and
  Sciancalepore}}]{el2018annealing}
\bibinfo{author}{\bibfnamefont{H.}~\bibnamefont{El~Dirani}},
  \bibinfo{author}{\bibfnamefont{A.}~\bibnamefont{Kamel}},
  \bibinfo{author}{\bibfnamefont{M.}~\bibnamefont{Casale}},
  \bibinfo{author}{\bibfnamefont{S.}~\bibnamefont{Kerdiles}},
  \bibinfo{author}{\bibfnamefont{C.}~\bibnamefont{Monat}},
  \bibinfo{author}{\bibfnamefont{X.}~\bibnamefont{Letartre}},
  \bibinfo{author}{\bibfnamefont{M.}~\bibnamefont{Pu}},
  \bibinfo{author}{\bibfnamefont{L.~K.} \bibnamefont{Oxenl{\o}we}},
  \bibinfo{author}{\bibfnamefont{K.}~\bibnamefont{Yvind}}, \bibnamefont{and}
  \bibinfo{author}{\bibfnamefont{C.}~\bibnamefont{Sciancalepore}},
  \bibinfo{journal}{Applied Physics Letters} \textbf{\bibinfo{volume}{113}},
  \bibinfo{pages}{081102} (\bibinfo{year}{2018}).

\bibitem[{\citenamefont{Gorodetsky et~al.}(2000)\citenamefont{Gorodetsky,
  Pryamikov, and Ilchenko}}]{gorodetsky2000rayleigh}
\bibinfo{author}{\bibfnamefont{M.~L.} \bibnamefont{Gorodetsky}},
  \bibinfo{author}{\bibfnamefont{A.~D.} \bibnamefont{Pryamikov}},
  \bibnamefont{and} \bibinfo{author}{\bibfnamefont{V.~S.}
  \bibnamefont{Ilchenko}}, \bibinfo{journal}{JOSA B}
  \textbf{\bibinfo{volume}{17}}, \bibinfo{pages}{1051} (\bibinfo{year}{2000}).

\bibitem[{\citenamefont{Zhao et~al.}(2020)\citenamefont{Zhao, Ji, Kim,
  Donvalka, Jang, Joshi, Yu, Joshi, Domeneguetti, Barbosa
  et~al.}}]{zhao2020visible}
\bibinfo{author}{\bibfnamefont{Y.}~\bibnamefont{Zhao}},
  \bibinfo{author}{\bibfnamefont{X.}~\bibnamefont{Ji}},
  \bibinfo{author}{\bibfnamefont{B.~Y.} \bibnamefont{Kim}},
  \bibinfo{author}{\bibfnamefont{P.~S.} \bibnamefont{Donvalka}},
  \bibinfo{author}{\bibfnamefont{J.~K.} \bibnamefont{Jang}},
  \bibinfo{author}{\bibfnamefont{C.}~\bibnamefont{Joshi}},
  \bibinfo{author}{\bibfnamefont{M.}~\bibnamefont{Yu}},
  \bibinfo{author}{\bibfnamefont{C.}~\bibnamefont{Joshi}},
  \bibinfo{author}{\bibfnamefont{R.~R.} \bibnamefont{Domeneguetti}},
  \bibinfo{author}{\bibfnamefont{F.~A.} \bibnamefont{Barbosa}},
  \bibnamefont{et~al.}, \bibinfo{journal}{Optica} \textbf{\bibinfo{volume}{7}},
  \bibinfo{pages}{135} (\bibinfo{year}{2020}).

\bibitem[{\citenamefont{Englund et~al.}(2005)\citenamefont{Englund, Fattal,
  Waks, Solomon, Zhang, Nakaoka, Arakawa, Yamamoto, and
  Vu{\v{c}}kovi{\'c}}}]{englund2005controlling}
\bibinfo{author}{\bibfnamefont{D.}~\bibnamefont{Englund}},
  \bibinfo{author}{\bibfnamefont{D.}~\bibnamefont{Fattal}},
  \bibinfo{author}{\bibfnamefont{E.}~\bibnamefont{Waks}},
  \bibinfo{author}{\bibfnamefont{G.}~\bibnamefont{Solomon}},
  \bibinfo{author}{\bibfnamefont{B.}~\bibnamefont{Zhang}},
  \bibinfo{author}{\bibfnamefont{T.}~\bibnamefont{Nakaoka}},
  \bibinfo{author}{\bibfnamefont{Y.}~\bibnamefont{Arakawa}},
  \bibinfo{author}{\bibfnamefont{Y.}~\bibnamefont{Yamamoto}}, \bibnamefont{and}
  \bibinfo{author}{\bibfnamefont{J.}~\bibnamefont{Vu{\v{c}}kovi{\'c}}},
  \bibinfo{journal}{Physical review letters} \textbf{\bibinfo{volume}{95}},
  \bibinfo{pages}{013904} (\bibinfo{year}{2005}).

\bibitem[{\citenamefont{Dieckmann et~al.}(1998)\citenamefont{Dieckmann,
  Spreeuw, Weidem{\"u}ller, and Walraven}}]{dieckmann1998two}
\bibinfo{author}{\bibfnamefont{K.}~\bibnamefont{Dieckmann}},
  \bibinfo{author}{\bibfnamefont{R.}~\bibnamefont{Spreeuw}},
  \bibinfo{author}{\bibfnamefont{M.}~\bibnamefont{Weidem{\"u}ller}},
  \bibnamefont{and} \bibinfo{author}{\bibfnamefont{J.}~\bibnamefont{Walraven}},
  \bibinfo{journal}{Physical Review A} \textbf{\bibinfo{volume}{58}},
  \bibinfo{pages}{3891} (\bibinfo{year}{1998}).

\bibitem[{\citenamefont{Thompson et~al.}(1992)\citenamefont{Thompson, Rempe,
  and Kimble}}]{thompson1992observation}
\bibinfo{author}{\bibfnamefont{R.}~\bibnamefont{Thompson}},
  \bibinfo{author}{\bibfnamefont{G.}~\bibnamefont{Rempe}}, \bibnamefont{and}
  \bibinfo{author}{\bibfnamefont{H.}~\bibnamefont{Kimble}},
  \bibinfo{journal}{Physical review letters} \textbf{\bibinfo{volume}{68}},
  \bibinfo{pages}{1132} (\bibinfo{year}{1992}).

\bibitem[{\citenamefont{Ramirez-Serrano
  et~al.}(2006)\citenamefont{Ramirez-Serrano, Yu, Kohel, Kellogg, and
  Maleki}}]{ramirez2006multistage}
\bibinfo{author}{\bibfnamefont{J.}~\bibnamefont{Ramirez-Serrano}},
  \bibinfo{author}{\bibfnamefont{N.}~\bibnamefont{Yu}},
  \bibinfo{author}{\bibfnamefont{J.~M.} \bibnamefont{Kohel}},
  \bibinfo{author}{\bibfnamefont{J.~R.} \bibnamefont{Kellogg}},
  \bibnamefont{and} \bibinfo{author}{\bibfnamefont{L.}~\bibnamefont{Maleki}},
  \bibinfo{journal}{Optics letters} \textbf{\bibinfo{volume}{31}},
  \bibinfo{pages}{682} (\bibinfo{year}{2006}).

\bibitem[{\citenamefont{Shomroni et~al.}(2014)\citenamefont{Shomroni,
  Rosenblum, Lovsky, Bechler, Guendelman, and Dayan}}]{shomroni2014all}
\bibinfo{author}{\bibfnamefont{I.}~\bibnamefont{Shomroni}},
  \bibinfo{author}{\bibfnamefont{S.}~\bibnamefont{Rosenblum}},
  \bibinfo{author}{\bibfnamefont{Y.}~\bibnamefont{Lovsky}},
  \bibinfo{author}{\bibfnamefont{O.}~\bibnamefont{Bechler}},
  \bibinfo{author}{\bibfnamefont{G.}~\bibnamefont{Guendelman}},
  \bibnamefont{and} \bibinfo{author}{\bibfnamefont{B.}~\bibnamefont{Dayan}},
  \bibinfo{journal}{Science} \textbf{\bibinfo{volume}{345}},
  \bibinfo{pages}{903} (\bibinfo{year}{2014}).

\bibitem[{\citenamefont{Alaeian et~al.}(2020)\citenamefont{Alaeian, Ritter,
  Basic, L{\"o}w, and Pfau}}]{alaeian2020cavity}
\bibinfo{author}{\bibfnamefont{H.}~\bibnamefont{Alaeian}},
  \bibinfo{author}{\bibfnamefont{R.}~\bibnamefont{Ritter}},
  \bibinfo{author}{\bibfnamefont{M.}~\bibnamefont{Basic}},
  \bibinfo{author}{\bibfnamefont{R.}~\bibnamefont{L{\"o}w}}, \bibnamefont{and}
  \bibinfo{author}{\bibfnamefont{T.}~\bibnamefont{Pfau}},
  \bibinfo{journal}{Applied Physics B} \textbf{\bibinfo{volume}{126}},
  \bibinfo{pages}{1} (\bibinfo{year}{2020}).

\end{thebibliography}


\end{document}